\documentclass[aps,prb,amsmath,twocolumn,amssymb,titlepage,showkeys,showpacs,floatfix,superscriptaddress,citeautoscript]{revtex4-1} 

\pdfoutput=1
\newcommand*{\romnum}[1]{\expandafter\@\romannumeral #1}
\usepackage{xcolor}

\usepackage{graphicx}

\makeatletter
\newcommand*{\rom}[1]{\expandafter\@slowromancap\romannumeral #1@}
\makeatother

\newcommand{\etal}{\mbox{\textit{et al.}}}
\newcommand{\Vs}{\ensuremath{V_{\rm S}}}
\newcommand{\Vd}{\ensuremath{V_{\rm D}}}
\newcommand{\Vsg}{\ensuremath{V_{\rm SG}}}
\newcommand{\Vbg}{\ensuremath{V_{\rm BG}}}
\newcommand{\Vsig}{\ensuremath{V_{\rm SiG}}}
\newcommand{\Vsd}{\ensuremath{V_{\rm SD}}}
\newcommand{\Iae}{\ensuremath{I_{\rm ae}}}

\begin{document}

\title{Quantised Charge Transport driven by a Surface Acoustic Wave in induced unipolar and bipolar junctions}

\author{Yousun \surname{Chung}}
\email{yousun.chung@unsw.edu.au}
\altaffiliation[Current Address: ]{Centre of Excellence for Quantum Computation and Communication Technology, University of New South Wales, Sydney, New South Wales 2052, Australia}
\affiliation{Cavendish Laboratory, University of Cambridge, Cambridge CB3 0HE, United Kingdom}

\author{Hangtian \surname{Hou}}
\affiliation{Cavendish Laboratory, University of Cambridge, Cambridge CB3 0HE, United Kingdom}

\author{Seok-Kyun \surname{Son}}
\altaffiliation[Current Address: ]{Department of Physics, Mokpo National University, Muan 58554, Republic of Korea}
\affiliation{Cavendish Laboratory, University of Cambridge, Cambridge CB3 0HE, United Kingdom}

\author{Tzu-Kan \surname{Hsiao}}
\altaffiliation[Current Address: ]{QuTech, Delft University of Technology, P.O. Box 5046, 2600 GA Delft, Netherlands}
\affiliation{Cavendish Laboratory, University of Cambridge, Cambridge CB3 0HE, United Kingdom}

\author{Ateeq \surname{Nasir}}
\affiliation{Cavendish Laboratory, University of Cambridge, Cambridge CB3 0HE, United Kingdom}
\affiliation{National Physical Laboratory, Hampton Road, Teddington, TW11 0LW, United Kingdom}

\author{Antonio \surname{Rubino}}
\affiliation{Cavendish Laboratory, University of Cambridge, Cambridge CB3 0HE, United Kingdom}

\author{Jonathan P. \surname{Griffiths}}
\affiliation{Cavendish Laboratory, University of Cambridge, Cambridge CB3 0HE, United Kingdom}

\author{Ian \surname{Farrer}}
\altaffiliation[Current Address: ]{Department of Electronic and Electrical Engineering, The University of Sheffield, Sheffield S1 3JD, United Kingdom}
\affiliation{Cavendish Laboratory, University of Cambridge, Cambridge CB3 0HE, United Kingdom}

\author{David A. \surname{Ritchie}}
\affiliation{Cavendish Laboratory, University of Cambridge, Cambridge CB3 0HE, United Kingdom}

\author{Christopher J. B. \surname{Ford}}
\email{cjbf@cam.ac.uk}
\affiliation{Cavendish Laboratory, University of Cambridge, Cambridge CB3 0HE, United Kingdom}


\begin{abstract}

Surface acoustic waves (SAWs) have been used to transport single electrons across long distances of several hundreds of microns. They can potentially be instrumental in the implementation of scalable quantum processors and quantum repeaters, by facilitating interaction between distant qubits. While most of the work thus far has focused on SAW devices in doped GaAs/AlGaAs heterostructures, we have developed a method of creating lateral \textit{p-n} junctions in an undoped heterostructure containing a quantum well, with the expected advantages of having reduced charge noise and increased spin-coherence lifetimes due to the lack of dopant scattering centres. We present experimental observations of SAW-driven single-electron quantised current in an undoped GaAs/AlGaAs heterostructure, where single electrons were transported between regions of induced electrons. We also demonstrate pumping of electrons by a SAW across the sub-micron depleted channel between regions of electrons and holes, and observe light emission at such a lateral \textit{p-n} junction. Improving the lateral confinement in the junction should make it possible to produce a quantised electron-to-hole current and hence SAW-driven emission of single photons.

\end{abstract}

\maketitle

For quantum cryptography and quantum information processing, a solid-state single-photon source capable of providing a controllable number of photons on demand with a high repetition rate is required.\cite{shields_2007} Foden \etal\cite{foden_2000} proposed that the stream of single electrons that can be driven by a surface acoustic wave (SAW) through a semiconductor channel could be directed into a sea of holes and hence generate single photons. The number of photons emitted should equal the number of electrons trapped in each SAW minimum, while the polarisation of the emitted photons depends on the spin state of the incoming carriers. It should therefore be possible to use SAWs to facilitate long-range interactions between qubits not only by physically transporting qubits in the SAW minima~\cite{delima_2004,barnes_2000,mcneil_2011b, hermelin_2011, aref_2015, schuetz_2015, bertrand2016fast, bertrand_2016}, but also by transferring spin information via SAW-generated polarised single photons\cite{kosaka_2009}.

We have previously shown that a SAW could carry a well-defined small number of electrons through a short pinched-off channel in a modulation-doped GaAs/AlGaAs heterostructure~\cite{shil_1996,talyanskii_1997}, by observing plateaux in the SAW-driven current as a function of the voltage on a split pair of gates on either side of the channel. The plateaux, of height $I = Nef$ (where $e$ is electronic charge and $f$ the SAW frequency), correspond to transport through the channel of an integer number $N$ of electrons trapped in each SAW minimum. $N$ is determined by the SAW potential and the maximum electrostatic potential gradient in the 1D channel. 

To emit single photons, such a SAW-driven stream of electrons must be dragged into a \textit{p}-type region for recombination with holes. However, fabricating lateral devices with both \textit{n}- and \textit{p}-type regions is non-trivial\cite{kim_1999,yuan_2002,hosey_2004,cecchini2005surface,smith_2006,kaestner_2006apl,gell_2007apl,dai_2014}. Electroluminescence (EL) has previously been observed from some lateral \textit{p-n} junctions\cite{yuan_2002,kaestner_2006apl,dai_2014}. SAW-driven light emission from a lateral \textit{p-n} junction has also been achieved, but this either required very complicated fabrication processes when using doped substrates (regrowth on patterned substrates) and wide channels~\cite{gell_2007apl}, or else electrons and holes were induced in undoped material by passing high currents within the separate regions, and there was no sign of quantisation of the \textit{n}-to-\textit{p} current.~\cite{simoni_2009} 

In this article, we present the first observation of SAW-induced quantised electron transport in an undoped system, where the use of realistic boundary conditions in self-consistent electrostatic potential calculations was vital in designing the device gate structure. We also demonstrate SAW-driven light emission from a \textit{p-n} junction formed by inducing controllable equilibrium populations of electrons and holes in close proximity in an undoped heterostructure using the same device design. We have thus developed the necessary ingredients for, and also demonstrated the first steps towards, a SAW-driven single-photon source.

\begin{figure}[t!]
{\centering\includegraphics[width=\linewidth]{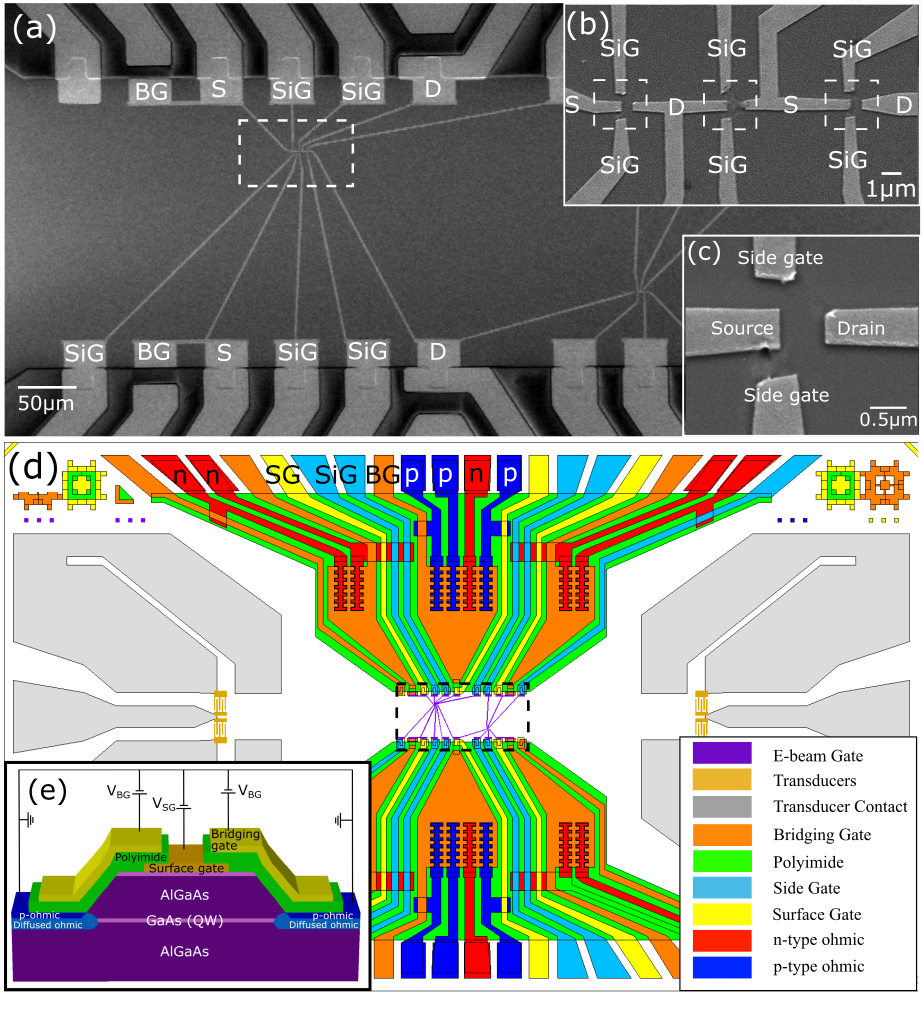}}
\caption{(a) Scanning-electron microscope (SEM) image of the active region of device A.
S, D, BG, and SiG are source and drain surface gates, bridging gates and side gates, respectively. Insulated BGs overlap the ohmic contacts, inducing carriers and leading them to the surface inducing gates. D can be used as an \textit{n}- or \textit{p}-type contact depending on the voltages applied to the corresponding bridging and surface gates. (b) SEM image of the region with three junctions. (c) One of the lateral \textit{n-i-n}, \textit{p-i-p}, or \textit{n-i-p} junctions. (d) Schematic of the whole device, where the corresponding contacts have been labelled. More details regarding device operation can be found in the supplementary material. (e) Cross-section of device, showing the quantum well and the gate stack.}
\label{fig1}
\end{figure}

Figure~\ref{fig1} shows the device layout, where depending on the applied voltages, electrons or holes can be induced, allowing permutations of lateral \textit{n-i-n}, \textit{p-i-p}, or \textit{n-i-p} junctions in the active region (figure~\ref{fig1}(b)). A single junction includes source and drain contacts (see figure~\ref{fig1}(c)), and two side gates to control the confinement of carriers in the 1D channel. We aim to have the length of the intrinsic region between source and drain ($\sim$600\,nm
) greater than half the SAW wavelength so that each SAW potential minimum becomes completely isolated from both leads at some point in the cycle. Using the results of our self-consistent electrostatic modelling, we chose the geometry and dimensions so that SAW transport across the junction in our devices is likely to be quantised by having a strongly confined SAW minimum. In order to generate SAWs, a high-frequency signal is applied to an interdigital transducer (IDT). The IDT periodicity of $1\mu$m gives a SAW resonant frequency of 2.848\,GHz.

\begin{figure}[t]
\includegraphics[width=\linewidth]{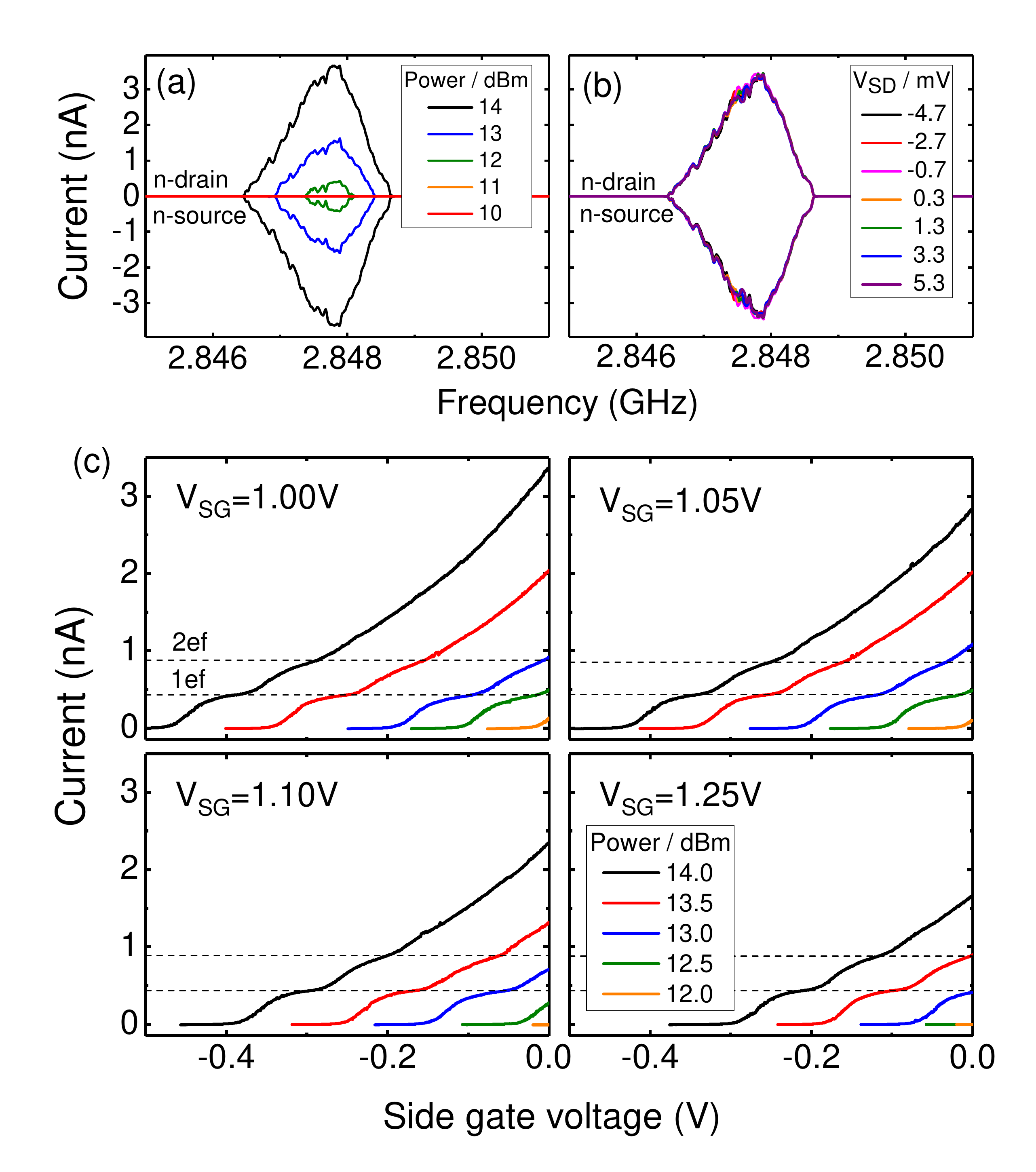}
\centering
\caption{The acoustoelectric current $\Iae$ of electrons driven by a SAW across an \textit{n-i-n} junction from S to D \textit{vs} SAW frequency $f$ (device A). D and S currents are equal and opposite. $\Vsg=1.1$\,V, $\Vsig=0$\,V, $T=4.2$\,K. (a) $I_{\rm ae}$ varying SAW power from 14 to 10\,dBm. (b) $\Iae$ varying S-D bias $\Vsd$ at 14\,dBm SAW power, $\Vbg=2.2$\,V, $\Vsg=1.15$\,V, $\Vsig=0$\,V. (c) $\Iae$ across the \textit{n-i-n} junction \textit{vs} side-gate voltage $\Vsig$ at different SAW powers and surface-gate voltages $\Vsg$, showing approximately quantised current plateaux close to multiples of $ef=456$\,pA ($f=2.848$\,GHz).}
\label{figIvsf_NN}
\end{figure}

For the \textit{n-i-n} junction devices discussed in this paper, the direction of the SAW is from the drain to the source, so that electrons are transported in this direction across the intrinsic region when the SAW amplitude is large enough that each SAW potential minimum retains a dip when added to the steep potential in the intrinsic region. Figure~\ref{figIvsf_NN} shows the SAW-driven electron current across an \textit{n-i-n} junction. Below 11\,dBm there was no current at the resonant frequency (see fig.~\ref{figIvsf_NN}a), implying that a SAW power of 11\,dBm was not enough to overcome the slope of the potential barrier in the intrinsic region and hence transport electrons across the junction. Above 11\,dBm, the current increased strongly with power. The source and drain currents were equal and opposite, showing that there was no leakage current. Note that for this measurement, there was no applied bias on the side gates, and also no external S-D bias, so the current was SAW-driven transport directly across the intrinsic region. This is further supported by bias-dependence measurements, which show that the current is independent of S-D bias in the range $\pm$5\,mV. This is consistent with SAW-driven transport without significant contributions from crosstalk or any conduction path through the side gates due to an external source-drain bias or side-gate voltage. 

Figure~\ref{figIvsf_NN}c shows experimental evidence of SAW quantisation, consistent with the results of the model, for voltages on the surface gates above the threshold at which carriers are induced under them. The measured current \textit{vs} side-gate bias shows quantised plateaux, as $N$ decreases, and eventually the SAW current goes to zero when the potential is too steep for the SAW to pump any electrons. The SAW power and surface gate (and hence the carrier density) were varied in order to try to optimise the confinement of the moving quantum dot, and hence increase the chance of observing quantisation, though little improvement was found. Two plateaux are visible, where the first plateau is found at 0.45\,nA and the second at 0.9\,nA, in agreement with expectations for a SAW-pumped single-electron current following the equation $I=Nef$. Probably partly owing to the measurement temperature used ($T=4.2$\,K), the plateaux are not as flat as can be achieved in doped devices \cite{ford_2017}. Nevertheless, our results show that it is possible to transfer a well-defined number $N$ of electrons in each SAW minimum and to control it using side-gate voltages. This is a significant step towards creating a single-photon source using an undoped induced \textit{n-i-p} junction device, in which a single electron would be transferred by the SAW in each cycle and would recombine with one of the holes induced on the other side of the junction. We present preliminary results on SAW-driven current and electroluminescence in the second part of this paper.

\begin{figure}[t]
\includegraphics[width=\linewidth]{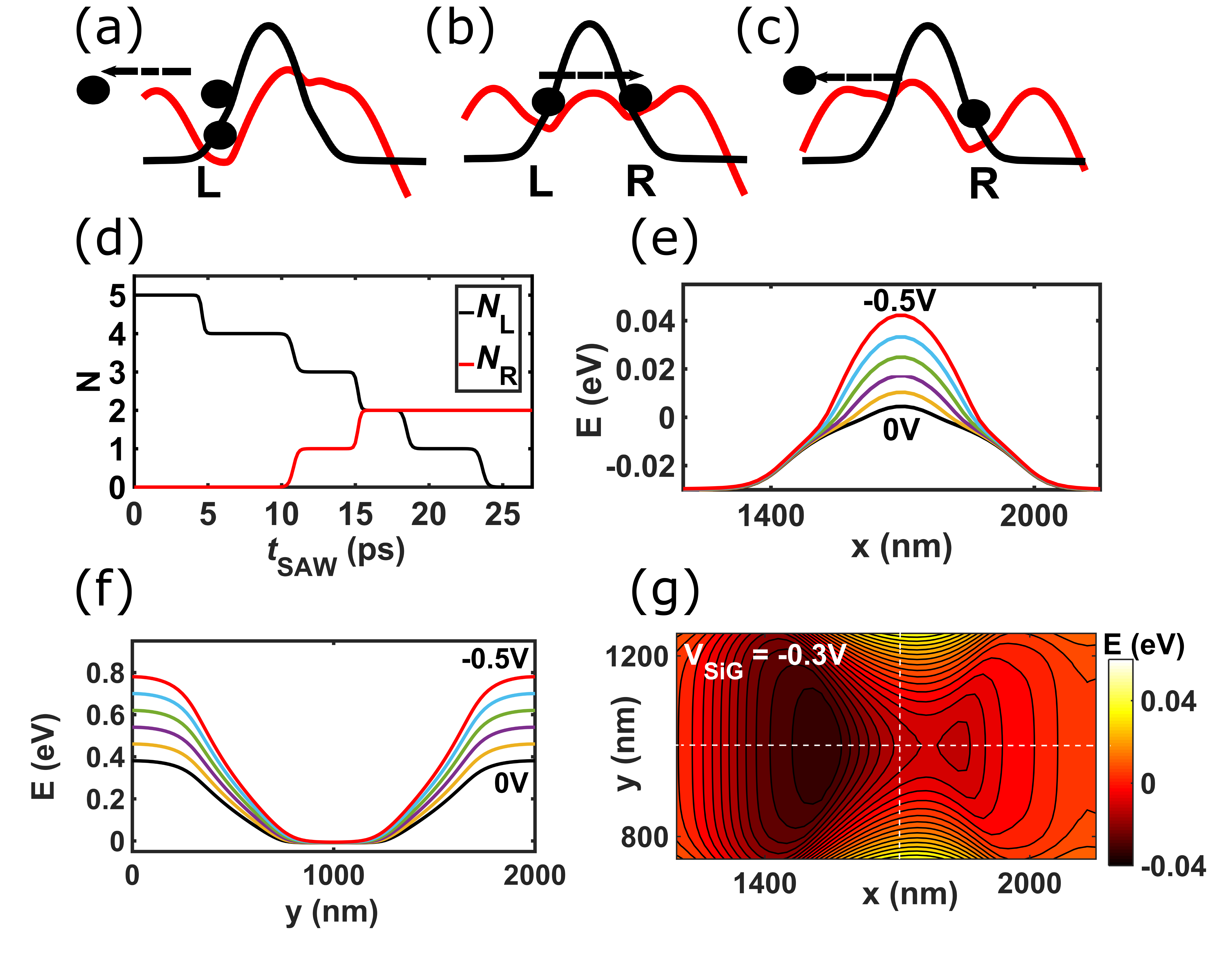}
\centering
\caption{(a--c) Model \textit{n-i-n} potential energy in the $x$-direction (black lines) and with an added SAW (red lines) at different times in a SAW cycle. Black dots are electrons trapped in potential minima, before and after the tunnelling events shown by arrows. (d) The number of electrons $N_{\rm L}$ and $N_{\rm R}$ in the two minima seen in (a--c) as a function of time $t_{\rm SAW}$ as the SAW travels through the junction. (e) Cut through the potential energy in the $y$-direction, along the white dashed line shown in (g) for $\Vsig$ from 0 to $-0.5$\,V in steps of 0.1\,V, as labelled. (f) Cut through the potential energy in the $x$-direction, along the white dashed line shown in (g) for $\Vsig$ from 0 to $-0.5$\,V in steps of 0.1\,V, as labelled. (g) The electrostatic potential energy in the $x-y$ plane at the depth of the QW in an \textit{n-i-n} junction at side-gate voltage $\Vsig=-0.3$\,V, with an added screened SAW potential of peak amplitude 25\,meV at $t_{\rm SAW}=10$\,ps.}

\label{simulation_NN}
\end{figure}

We now discuss the effect of gate geometry on the quantisation of the SAW-driven current. In developing an electrostatic model for undoped GaAs devices, it is important to note that, unlike in doped devices, assigning the correct boundary conditions is critical and non-trivial, because of the absence of modulation doping, and hence of built-in electric fields. In a doped GaAs device it is common to assume that the Fermi energy is pinned at the regrowth interface $\sim 1\,\mu$m below the surface. By comparing the model and experiments, we have shown that this boundary condition is inappropriate as charges in dangling bonds there become frozen at cryogenic temperatures, and a deep boundary condition in the GaAs substrate gives more accurate results, and indeed we have shown that it is essential in calculations for an undoped device.\cite{hou_2018} Figure~\ref{simulation_NN}(g) shows the calculated electrostatic potential energy in an induced \textit{n-i-n} junction matching the experimental design. The junction entrance presents a potential slope of $\sim20$--30\,meV/$\mu$m, 
which the SAW is expected to be able to overcome with applied SAW power above $10-15$\,dBm \cite{schneble_2006,kataoka_2006}, comparable to our experimental results in figure~\ref{figIvsf_NN}. In the experiment, the side-gate pinch-off voltage becomes more negative with increasing SAW power because the higher SAW amplitude enables pumping up the steeper slope. This voltage ranges from 0 to $-0.4$\,V for powers ranging from 12 to 14\,dBm, in agreement with the simulation results. As the side-gate voltage becomes more negative, the transverse confinement becomes stronger and this, together with the SAW longitudinal confinement, defines a dynamic quantum dot. For a given voltage there is some $N$ for which, with $(N+1)$ electrons in the dynamic quantum dot, Coulomb charging of the dot brings the highest-energy electron's energy close to the top of the back tunnel barrier and so it is highly likely that the electron tunnels or falls out of the dot back into the original lead, leaving $N$ electrons in the dot to reach the other lead, and this produces a quantised SAW-driven current. We estimate the charging energy to be $\sim 3$\,meV from our previous work \cite{astley_2007}.

We use a simple model to calculate the acoustoelectric current through the channel, treating the potential barriers before and after the dot as saddle points with known tunnelling-probability functions.\cite{buttiker_1990} Figure~\ref{simulation_NN}(a--c) shows the calculated potential without (black) and with (red) an applied SAW potential (partially screened in the leads), at various stages of the SAW propagation. Firstly, many electrons collect in the minimum before the barrier (labelled L in (a)), and as the SAW propagates towards the intrinsic barrier, the minimum starts to shrink and some electrons tunnel back into the source (a). As the SAW continues propagating, the intrinsic barrier becomes lower than the following SAW maximum (b) so electrons start to tunnel across the barrier and reach the minimum on the right (R) and hence the right lead. Later the intrinsic barrier becomes high again, and electrons remaining in minimum L go back to the source again (c). Figure~\ref{simulation_NN}(d) shows the calculated number of electrons $N_{\rm L}$ and $N_{\rm R}$ in minima L and R seen in (a--c) as a function of time as one wavelength of the SAW travels through the junction. In this case, $N_{\rm R}=2$ at the end of the cycle, resulting in a quantised acoustoelectric current $I=2ef$.

We now turn our attention to device operation with different types of carrier induced on each side (forming an \textit{n-i-p} junction). As a precursor to a SAW-driven single-photon source, we demonstrate SAW-driven electroluminescence (EL) generated by 
electrons (holes) from an \textit{n}-type (\textit{p}-type) region transported by the SAW across the intrinsic region recombining with the induced holes (electrons) in the \textit{p}-type (\textit{n}-type) region there. As the chemical potentials in the S and D leads lie in different bands (valence and conduction), there is a repulsive potential slope in between S and D. One can forward-bias the junction close to the flat-band condition to drive a current and this is likely to give photon emission near the junction even without a SAW, but there is no reason for the photons to come out singly. It should be possible to use SAWs to transport carriers across the intrinsic region with a smaller S-D bias (less than the band gap), just high enough to help the SAW overcome the potential slope between S and D but not enough to transport charge without a SAW. Then, if the channel is pinched off enough to allow only one electron (hole) per SAW minimum, this single charge will recombine to produce a single photon, ideally before another charge arrives in the next cycle.

\begin{figure}[tp]
\includegraphics[width=\linewidth]{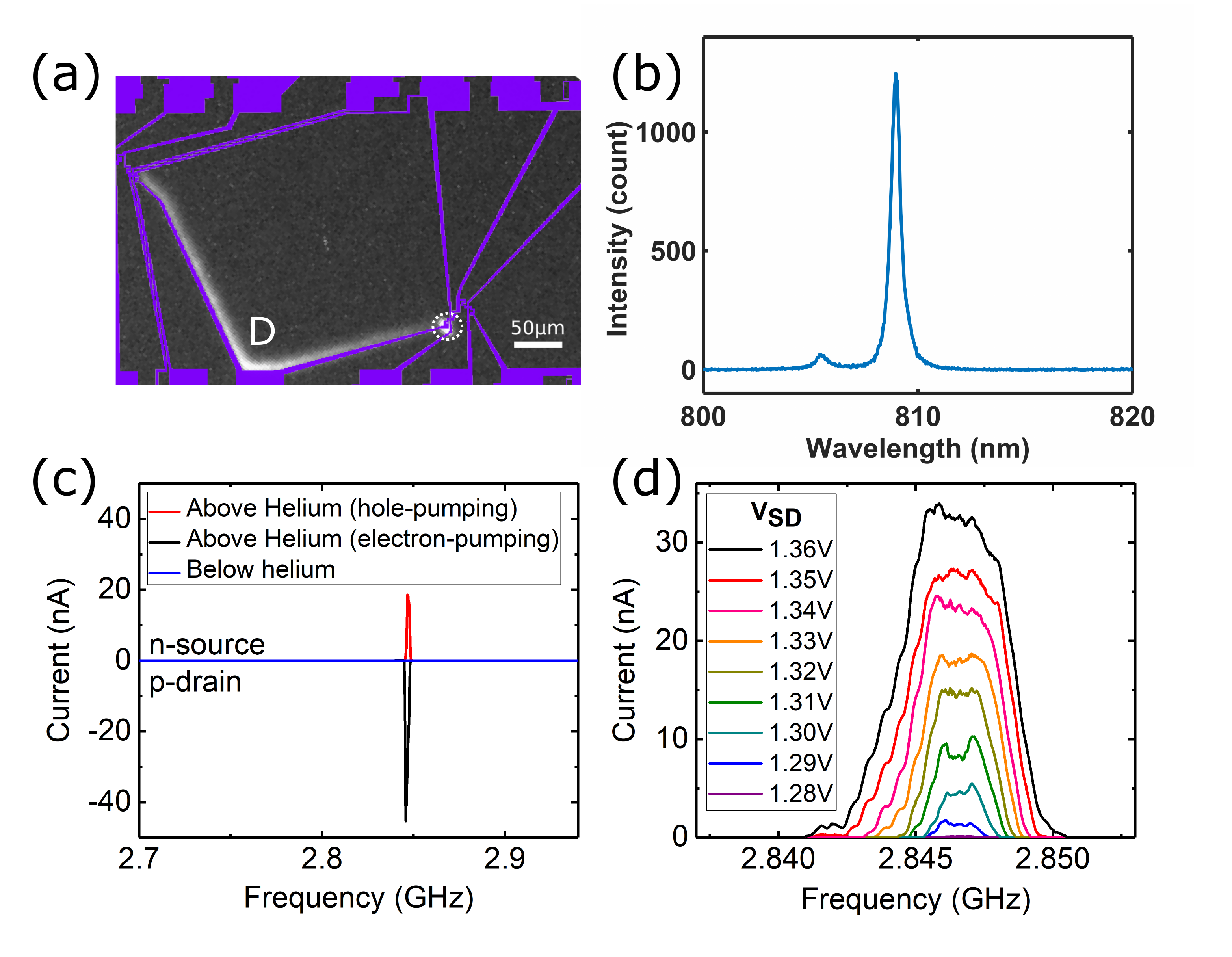}
\centering
\caption{(a) EL image of device B without SAW at $\Vsig=0$\,V, ${\Vsg}_{\rm source}=2$\,V, ${\Vsg}_{\rm drain}=-2$\,V, and $\Vsd=2.5$\,V, where most of the EL was observed at the \textit{p-n} junction region, indicated by the dashed circle. The gate design is overlaid (in purple online), aligned from a later optical image. (b) Spectrum taken from the \textit{p-n} junction enclosed by dashed circle in (a) ($\Vsig=0$\,V, ${\Vsg}_{\rm source}=2.5$\,V, ${\Vsg}_{\rm drain}=-2.27$\,V, $\Vsd=3.1$\,V). 
(c) SAW-driven current of electrons (holes) across an \textit{n-i-p} junction from the \textit{n}-type source (\textit{p}-type drain) \textit{vs} $f$ (device A, $\Vsd=1.4$\,V SAW power 14\,dBm, $\Vsig=1.08$\,V) when device above liquid-helium surface (no damping). Current disappears on immersion in liquid, which damps the SAW. (d) SAW-driven current for varying S-D bias $\Vsd$ (other conditions are as in (c)).}
\label{figIvsf_NP}
\end{figure}

EL measurements (see Fig.~\ref{figIvsf_NP}a and b) without a SAW were performed on a second device (B, where some shallow etching was used to confine electrons near to the gates, and using a wafer with a smaller QW width), as a means of confirming that the \textit{p}- and \textit{n}-type regions were induced at the correct interface. Photon emission was observed under a high (2.5\,V) applied bias. The strongest EL is observed from a 6--10\,$\mu$m-diameter region at the junction, though the resolution of the image is low so the emission spot may actually be smaller than this. The additional EL around the D surface gate (labelled 'D' in fig.~\ref{figIvsf_NP}a) is likely to be due to a lack of confinement for electrons once they reach the region of holes: under such a high S-D bias, electrons with a high kinetic energy can leave the junction region without recombination and eventually recombine with holes under the D surface gate, causing the L-shaped EL (an etched channel alongside this gate prevents the electrons escaping into the central intrinsic region). The measurement was performed at $\sim 20$\,K. The spectrum obtained at the \textit{p-n} junction (Fig.~\ref{figIvsf_NP}b) has a peak at a wavelength of about 809\,nm, as expected for the quantum well in this wafer. 

Fig.~\ref{figIvsf_NP}c shows the SAW-driven current of electrons or holes across the same \textit{n-i-p} junction using different transducers at either end of the junction. With a positive voltage on the source BG ($\Vbg=2.2$\,V), $\Vsg=1.15$\,V was applied to the corresponding surface gate in order to induce electrons at the source near the central active region. Similarly, with a negative voltage on the drain BG ($\Vbg=-2.0$\,V), $\Vsg=-1.05$\,V was applied to the corresponding surface gate in order to induce holes at the drain. A bias of 1.4\,V was applied to the drain relative to the grounded source ($\Vsd=1.4$\,V). This is just below the threshold bias (at which the applied bias becomes greater than the band gap ($\sim 1.52$\,eV), causing current flow across the intrinsic region). The height of the potential barrier at the intrinsic region is then low enough to enable a SAW coming from S to transfer electrons across the junction from S (\textit{n}) to D (\textit{p}) or a SAW from the other IDT to carry holes the other way. When pumping electrons, a side-gate voltage $\Vsig = 1.0$\,V was applied relative to S  ($-0.4$\,V relative to D) to attract electrons and help extract them from S, so that when a SAW is applied using the IDT at the source (\textit{n}) end, it can transport electrons to D. Hence, at the resonant frequency of 2.846\,GHz, there was a large peak in acoustoelectric current across the intrinsic region between S and D.

Under the same source-drain bias, when the SAW direction was instead from D to S and $\Vsig=0.5$\,V ($-0.9$\,V relative to D (\textit{p})), holes were pumped from D to S. To confirm that the current was driven by the SAW, the device was immersed in liquid helium and the current was observed to drop to zero, as expected when the SAW is damped by the liquid. Any current driven by the electromagnetic wave itself (crosstalk) would be unaffected. We can rule out the possibility that the current flows because at some stage in the SAW cycle the junction potential is pulled entirely below the Fermi energy by a SAW minimum. This could occur for an \textit{n-i-n} junction but not for an \textit{n-i-p} junction, as in the latter case energy is also needed to reach the far side. Also, the SAW half wavelength was less than the intrinsic region's length. No current quantisation was seen like that in Figure~\ref{figIvsf_NN} for the \textit{n-i-n} junction, but it should be possible to quantise SAW transport across the junction if stronger lateral confinement can be obtained to provide a larger charging energy in future experiments \cite{kataoka_2006}.

\begin{figure}[!t]
\includegraphics[width=0.7\linewidth]{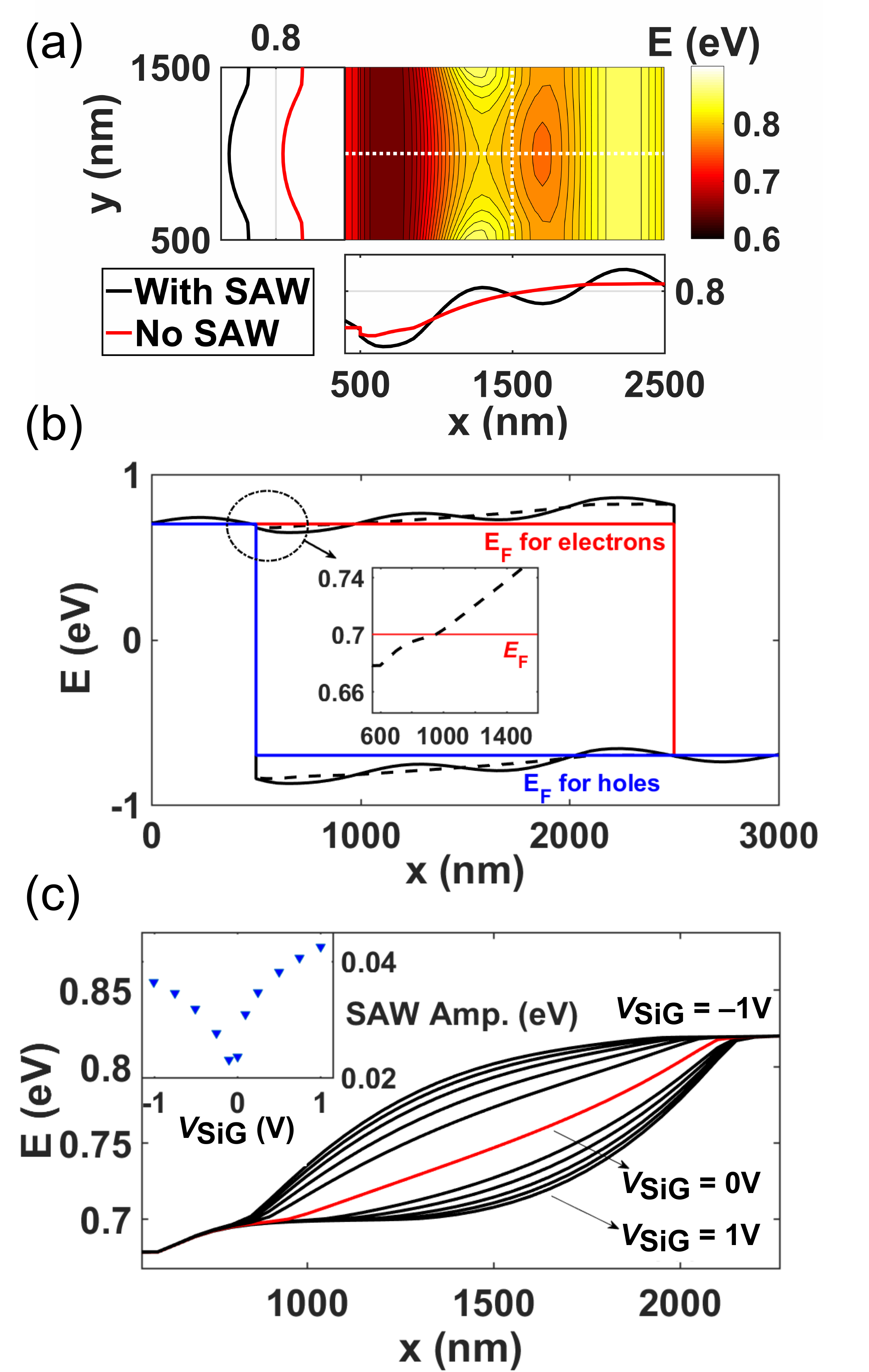}
\centering
\caption{Modelling an \textit{n-i-p} junction. (a) 2D electrostatic energy landscape in the $x-y$ plane with added SAW potential ($\Vsig=0$\,V, ${\Vsg}_{\rm source}=1$\,V, ${\Vsg}_{\rm drain}=-1$\,V). Cross-sections in the $x$- and $y$-directions along the white dotted lines are also shown with and without SAW. (b) Energy band-structure with added SAW (source bias $\Vs=-0.7$\,V, drain bias $\Vd=0.7$\,V). Inset shows enlargement of conduction band (dashed line) when electrons induced below Fermi level. Fermi level is $0.7$\,V for electrons (red) and $-0.7$\,V for holes (blue). (c) Conduction-band profile through junction with different $\Vsig$ (${\Vsg}_{\rm source}=1$\,V, ${\Vsg}_{\rm drain}=-1$\,V, $\Vs=-0.7$\,V, $\Vd=0.7$\,V). Inset: required SAW amplitude \textit{vs} $\Vsig$.}
\label{simulation_NP}
\end{figure}

Figure~\ref{figIvsf_NP}d shows the current across the \textit{n-i-p} junction as a function of SAW frequency as the applied bias on the drain is stepped from 1.28\,V to 1.36\,V in 10\,mV steps with a slow sweep rate. The measurement was performed in such a way that the bias was incremented non-monotonically so that the time-dependent drift could be shown to be negligible. A voltage of $\Vsig=1.08$\,V was applied to the side gates to assist the SAW to drag electrons from the source to the region of holes at the drain. It is observed that the SAW-driven current increases with source-drain bias. This is expected as the source-drain bias creates an electric field that lowers the potential barrier in the intrinsic region, allowing the SAW to carry more electrons in each minimum. The width of the current peak in frequency also becomes wider as the source-drain bias is increased, unlike that observed for the \textit{n-i-n} or \textit{p-i-p} junctions. The resonant peak in the \textit{n-i-p} case is broadened to a width of 1\,MHz (from 2.846 to 2.847\,GHz), still less than the pass-band of the IDT, as more of the tails of the SAW resonant peak exceeded the pumping threshold amplitude.

We have simulated the electrostatic potential distribution of our \textit{n-i-p} junction using the same dimensions and voltages as for device A. Figure~\ref{simulation_NP}(a) shows conduction and valence bands (dashed lines) at the junction. An external electric field applied by a surface inducing gate pulls the conduction (valence) band below the appropriate Fermi level, inducing free electrons (holes) in the QW. Figure~\ref{simulation_NP}(b) illustrates how the slope of the potential barrier in the intrinsic junction region changes with and without a SAW. Applying a non-zero side-gate voltage increases the slope of the potential barrier (figure~\ref{simulation_NP}(c)), so the required SAW amplitude increases towards $\sim 40$\,meV for $\Vsig \sim \pm$ 1V (see inset), which is achievable with the current experimental set-up. However, because the side-gate voltage required to squeeze one species of carrier attracts the other species, it is necessary to implement a way of reducing unwanted charge carriers in the channel, and hence increase the probability of recombination between electrons and holes at the junction. That is why Device B and subsequent devices have an additional high-resolution etched trench on either side of the channel, under the e-beam-defined side gates and around the surface gates. Thus, when induced charges are transferred from the source or drain, they should be trapped near the junction before eventually escaping towards the leads, increasing the probability of recombination around the \textit{n-i-p} junction and consequently allowing photon emission to be more readily detectable there. Etching the regions under the side gates removes any alternative current paths. Such an etched device continued to show acoustoelectric current for several days longer than unetched devices, which lasted only a day or so before accumulation of charge in intrinsic regions or at the polyimide insulator interface made current flow impossible \cite{hsiao_2019}. An additional step that can be taken to further increase the lateral confinement in the channel is to add another pair of split side gates in order to allow extra control of the potential along the channel.

In summary, using gates to induce high-mobility 2D electron and hole gases in close proximity, we have fabricated lateral \textit{n-i-n} and \textit{n-i-p} junctions, through which a SAW can pump electrons or holes. We observed quantisation of this current in an \textit{n-i-n} junction. Changing the inducing voltages on one of the regions allows SAW-driven recombination of the pumped charges, from which we observe electroluminescence. In principle, this design can also be implemented in other semiconducting 2D systems, such as MoS$_2$ by placing them on a piezoelectric substrate. This demonstrates the potential development of this device architecture for a versatile, all-solid-state, single-photon source with high repetition rate only limited by emission and collection efficiency and the SAW frequency.

\textit{Acknowledgments.} We are grateful for useful discussions with Dr.\ Stuart Holmes, Prof.\ Richard Phillips and Prof.\ Crispin Barnes.
\bibliography{Quantised_SAW}

\end{document}